
%
%
\expandafter\ifx\csname phyzzx\endcsname\relax\else
 \errhelp{Hit <CR> and go ahead.}
 \errmessage{PHYZZX macros are already loaded or input. }
 \endinput \fi
\catcode`\@=11 
%
%
%

\font\fourteenrm=cmr10 scaled\magstep2
\font\twelverm=cmr10 scaled\magstep1
\font\ninerm=cmr9            \font\sixrm=cmr6

\font\fourteenbf=cmbx10 scaled\magstep2
\font\twelvebf=cmbx10 scaled\magstep1
\font\ninebf=cmbx9            \font\sixbf=cmbx6
\font\seventeeni=cmmi10 scaled\magstep3     \skewchar\seventeeni='177
\font\fourteeni=cmmi10 scaled\magstep2      \skewchar\fourteeni='177
\font\twelvei=cmmi10 scaled\magstep1        \skewchar\twelvei='177
\font\ninei=cmmi9                           \skewchar\ninei='177
\font\sixi=cmmi6                            \skewchar\sixi='177
\font\seventeensy=cmsy10 scaled\magstep3    \skewchar\seventeensy='60
\font\fourteensy=cmsy10 scaled\magstep2     \skewchar\fourteensy='60
\font\twelvesy=cmsy10 scaled\magstep1       \skewchar\twelvesy='60
\font\ninesy=cmsy9                          \skewchar\ninesy='60
\font\sixsy=cmsy6                           \skewchar\sixsy='60

\font\fourteenex=cmex10 scaled\magstep2
\font\twelveex=cmex10 scaled\magstep1
%

\font\fourteensl=cmsl10 scaled\magstep2
\font\twelvesl=cmsl10 scaled\magstep1
\font\ninesl=cmsl9

\font\fourteenit=cmti10 scaled\magstep2
\font\twelveit=cmti10 scaled\magstep1
\font\nineit=cmti9
\font\fourteentt=cmtt10 scaled\magstep2
\font\twelvett=cmtt10 scaled\magstep1
\font\fourteencp=cmcsc10 scaled\magstep2
\font\twelvecp=cmcsc10 scaled\magstep1
\font\tencp=cmcsc10
\newfam\cpfam
\newdimen\b@gheight		\b@gheight=12pt
\newcount\f@ntkey		\f@ntkey=0
\def\f@m{\afterassignment\samef@nt\f@ntkey=}
\def\samef@nt{\fam=\f@ntkey \the\textfont\f@ntkey\relax}
\def\rm{\f@m0 }
\def\mit{\f@m1 }         
\def\cal{\f@m2 }
\def\it{\f@m\itfam}
\def\sl{\f@m\slfam}
\def\bf{\f@m\bffam}
\def\tt{\f@m\ttfam}
\def\caps{\f@m\cpfam}
\def\fourteenpoint{\relax
    \textfont0=\fourteenrm          \scriptfont0=\tenrm
      \scriptscriptfont0=\sevenrm
    \textfont1=\fourteeni           \scriptfont1=\teni
      \scriptscriptfont1=\seveni
    \textfont2=\fourteensy          \scriptfont2=\tensy
      \scriptscriptfont2=\sevensy
    \textfont3=\fourteenex          \scriptfont3=\twelveex
      \scriptscriptfont3=\tenex
    \textfont\itfam=\fourteenit     \scriptfont\itfam=\tenit
    \textfont\slfam=\fourteensl     \scriptfont\slfam=\tensl
    \textfont\bffam=\fourteenbf     \scriptfont\bffam=\tenbf
      \scriptscriptfont\bffam=\sevenbf
    \textfont\ttfam=\fourteentt
    \textfont\cpfam=\fourteencp
    \samef@nt
    \b@gheight=14pt
    \setbox\strutbox=\hbox{\vrule height 0.85\b@gheight
				depth 0.35\b@gheight width\z@ }}
\def\twelvepoint{\relax
    \textfont0=\twelverm          \scriptfont0=\ninerm
      \scriptscriptfont0=\sixrm
    \textfont1=\twelvei           \scriptfont1=\ninei
      \scriptscriptfont1=\sixi
    \textfont2=\twelvesy           \scriptfont2=\ninesy
      \scriptscriptfont2=\sixsy
    \textfont3=\twelveex          \scriptfont3=\tenex
      \scriptscriptfont3=\tenex
    \textfont\itfam=\twelveit     \scriptfont\itfam=\nineit
    \textfont\slfam=\twelvesl     \scriptfont\slfam=\ninesl
    \textfont\bffam=\twelvebf     \scriptfont\bffam=\ninebf
      \scriptscriptfont\bffam=\sixbf
    \textfont\ttfam=\twelvett
    \textfont\cpfam=\twelvecp
    \samef@nt
    \b@gheight=12pt
    \setbox\strutbox=\hbox{\vrule height 0.85\b@gheight
				depth 0.35\b@gheight width\z@ }}
\def\tenpoint{\relax
    \textfont0=\tenrm          \scriptfont0=\sevenrm
      \scriptscriptfont0=\fiverm
    \textfont1=\teni           \scriptfont1=\seveni
      \scriptscriptfont1=\fivei
    \textfont2=\tensy          \scriptfont2=\sevensy
      \scriptscriptfont2=\fivesy
    \textfont3=\tenex          \scriptfont3=\tenex
      \scriptscriptfont3=\tenex
    \textfont\itfam=\tenit     \scriptfont\itfam=\seveni
    \textfont\slfam=\tensl     \scriptfont\slfam=\sevenrm
    \textfont\bffam=\tenbf     \scriptfont\bffam=\sevenbf
      \scriptscriptfont\bffam=\fivebf
    \textfont\ttfam=\tentt
    \textfont\cpfam=\tencp
    \samef@nt
    \b@gheight=10pt
    \setbox\strutbox=\hbox{\vrule height 0.85\b@gheight
				depth 0.35\b@gheight width\z@ }}
%
%
%
\normalbaselineskip = 20pt plus 0.2pt minus 0.1pt
\normallineskip = 1.5pt plus 0.1pt minus 0.1pt
\normallineskiplimit = 1.5pt
\newskip\normaldisplayskip
\normaldisplayskip = 20pt plus 5pt minus 10pt
\newskip\normaldispshortskip
\normaldispshortskip = 6pt plus 5pt
\newskip\normalparskip
\normalparskip = 6pt plus 2pt minus 1pt
\newskip\skipregister
\skipregister = 5pt plus 2pt minus 1.5pt
\newif\ifsingl@    \newif\ifdoubl@
\newif\iftwelv@    \twelv@true
\def\singlespace{\singl@true\doubl@false\spaces@t}
\def\doublespace{\singl@false\doubl@true\spaces@t}
\def\normalspace{\baselineskip=20pt}
\def\Tenpoint{\tenpoint\twelv@false\spaces@t}
\def\Twelvepoint{\twelvepoint\twelv@true\spaces@t}
\def\spaces@t{\relax
      \iftwelv@ \ifsingl@\subspaces@t3:4;\else\subspaces@t1:1;\fi
       \else \ifsingl@\subspaces@t3:5;\else\subspaces@t4:5;\fi \fi
      \ifdoubl@ \multiply\baselineskip by 5
         \divide\baselineskip by 4 \fi }
\def\subspaces@t#1:#2;{
      \baselineskip = \normalbaselineskip
      \multiply\baselineskip by #1 \divide\baselineskip by #2
      \lineskip = \normallineskip
      \multiply\lineskip by #1 \divide\lineskip by #2
      \lineskiplimit = \normallineskiplimit
      \multiply\lineskiplimit by #1 \divide\lineskiplimit by #2
      \parskip = \normalparskip
      \multiply\parskip by #1 \divide\parskip by #2
      \abovedisplayskip = \normaldisplayskip
      \multiply\abovedisplayskip by #1 \divide\abovedisplayskip by #2
      \belowdisplayskip = \abovedisplayskip
      \abovedisplayshortskip = \normaldispshortskip
      \multiply\abovedisplayshortskip by #1
        \divide\abovedisplayshortskip by #2
      \belowdisplayshortskip = \abovedisplayshortskip
      \advance\belowdisplayshortskip by \belowdisplayskip
      \divide\belowdisplayshortskip by 2
      \smallskipamount = \skipregister
      \multiply\smallskipamount by #1 \divide\smallskipamount by #2
      \medskipamount = \smallskipamount \multiply\medskipamount by 2
      \bigskipamount = \smallskipamount \multiply\bigskipamount by 4 }
\def\normalbaselines{ \baselineskip=\normalbaselineskip
   \lineskip=\normallineskip \lineskiplimit=\normallineskip
   \iftwelv@\else \multiply\baselineskip by 4 \divide\baselineskip by 5
     \multiply\lineskiplimit by 4 \divide\lineskiplimit by 5
     \multiply\lineskip by 4 \divide\lineskip by 5 \fi }
%
%
\interlinepenalty=50
\interfootnotelinepenalty=5000
\predisplaypenalty=9000
\postdisplaypenalty=500
\hfuzz=1pt
\vfuzz=0.2pt
\voffset=0pt
\dimen\footins=8 truein
%
%
%
\def\pagecontents{
   \ifvoid\topins\else\unvbox\topins\vskip\skip\topins\fi
   \dimen@ = \dp255 \unvbox255
   \ifvoid\footins\else\vskip\skip\footins\footrule\unvbox\footins\fi
   \ifr@ggedbottom \kern-\dimen@ \vfil \fi }
\def\makeheadline{\vbox to 0pt{ \skip@=\topskip
      \advance\skip@ by -12pt \advance\skip@ by -2\normalbaselineskip
      \vskip\skip@ \line{\vbox to 12pt{}\the\headline} \vss
      }\nointerlineskip}
\def\makefootline{\baselineskip = 1.5\normalbaselineskip
                 \line{\the\footline}}
\newif\iffrontpage
\newif\ifletterstyle
\newif\ifp@genum
\def\nopagenumbers{\p@genumfalse}
\def\pagenumbers{\p@genumtrue}
\pagenumbers
\newtoks\paperheadline
\newtoks\letterheadline
\newtoks\paperfootline
\newtoks\letterfootline
\newtoks\letterinfo
\newtoks\Letterinfo
\newtoks\date
\footline={\ifletterstyle\the\letterfootline\else\the\paperfootline\fi}
\paperfootline={\hss\iffrontpage\else\ifp@genum\tenrm\folio\hss\fi\fi}
\letterfootline={\iffrontpage\the\letterinfo\else\hfil\fi}
\Letterinfo={\hfil}
\letterinfo={\hfil}
\headline={\ifletterstyle\the\letterheadline\else\the\paperheadline\fi}
\paperheadline={\hfil}
\letterheadline{\iffrontpage \hfil \else
    \rm \ifp@genum page \ \folio\fi \hfil\the\date \fi \cr\hfil\the\pubnu \fi}
\def\monthname{\relax\ifcase\month 0/\or January\or February\or
   March\or April\or May\or June\or July\or August\or September\or
   October\or November\or December\else\number\month/\fi}
\def\today{\monthname\ \number\day, \number\year}
\date={\today}
\countdef\pageno=1      \countdef\pagen@=0
\countdef\pagenumber=1  \pagenumber=1
\def\advancepageno{\global\advance\pagen@ by 1
   \ifnum\pagenumber<0 \global\advance\pagenumber by -1
    \else\global\advance\pagenumber by 1 \fi \global\frontpagefalse }
\def\folio{\ifnum\pagenumber<0 \romannumeral-\pagenumber
           \else \number\pagenumber \fi }
\def\footrule{\dimen@=\prevdepth\nointerlineskip
   \vbox to 0pt{\vskip -0.25\baselineskip \hrule width 0.35\hsize \vss}
   \prevdepth=\dimen@ }
\newtoks\foottokens
\foottokens={}
\newdimen\footindent
\footindent=24pt
\def\vfootnote#1{\insert\footins\bgroup
   \interlinepenalty=\interfootnotelinepenalty \floatingpenalty=20000
   \singl@true\doubl@false\Tenpoint
   \splittopskip=\ht\strutbox \boxmaxdepth=\dp\strutbox
   \leftskip=\footindent \rightskip=\z@skip
   \parindent=1.2truecm
   \spaceskip=\z@skip \xspaceskip=\z@skip
   \the\foottokens
   \Textindent{$ #1 $}\footstrut\futurelet\next\fo@t}
\def\Textindent#1{\noindent\llap{#1\enspace}\ignorespaces}
\def\footnote#1{\attach{#1}\vfootnote{#1}}

\let\footsymbol=\star
\newcount\lastf@@t           \lastf@@t=-1
\newcount\footsymbolcount    \footsymbolcount=0
\newif\ifPhysRev
\def\bumpfootsymbolcount{\relax
   \iffrontpage \bumpfootsymbolNP \else \advance\lastf@@t by 1
     \ifPhysRev \bumpfootsymbolPR \else \bumpfootsymbolNP \fi \fi
   \global\lastf@@t=\pagen@ }
\def\bumpfootsymbolNP{\ifnum\footsymbolcount <0 \global\footsymbolcount =0 \fi
    \ifnum\lastf@@t<\pagen@ \global\footsymbolcount=0
     \else \global\advance\footsymbolcount by 1 \fi }
\def\bumpfootsymbolPR{\ifnum\footsymbolcount >0 \global\footsymbolcount =0 \fi
      \global\advance\footsymbolcount by -1 }
\def\fd@f#1 {\xdef\footsymbol{\mathchar"#1 }}
\def\generatefootsymbol{\ifcase\footsymbolcount \fd@f 13F \or \fd@f 279
	\or \fd@f 27A \or \fd@f 278 \or \fd@f 27B \else
	\ifnum\footsymbolcount <0 \fd@f{023 \number-\footsymbolcount }
	 \else \fd@f 203 {\loop \ifnum\footsymbolcount >5
		\fd@f{203 \footsymbol } \advance\footsymbolcount by -1
		\repeat }\fi \fi }

\def\nonfrenchspacing{\sfcode`\.=3001 \sfcode`\!=3000 \sfcode`\?=3000
	\sfcode`\:=2000 \sfcode`\;=1500 \sfcode`\,=1251 }
\nonfrenchspacing
\newdimen\d@twidth
{\setbox0=\hbox{s.} \global\d@twidth=\wd0 \setbox0=\hbox{s}
	\global\advance\d@twidth by -\wd0 }
\def\removehglue{\loop \unskip \ifdim\lastskip >\z@ \repeat }
\def\roll@ver#1{\removehglue \nobreak \count255 =\spacefactor \dimen@=\z@
	\ifnum\count255 =3001 \dimen@=\d@twidth \fi
	\ifnum\count255 =1251 \dimen@=\d@twidth \fi
    \iftwelv@ \kern-\dimen@ \else \kern-0.83\dimen@ \fi
   #1\spacefactor=\count255 }
\def\step@ver#1{\relax \ifmmode #1\else \ifhmode
	\roll@ver{${}#1$}\else {\setbox0=\hbox{${}#1$}}\fi\fi }
\def\attach#1{\step@ver{\strut^{\mkern 2mu #1} }}
%
%
%
\newcount\chapternumber      \chapternumber=0
\newcount\sectionnumber      \sectionnumber=0
\newcount\equanumber         \equanumber=0
\let\chapterlabel=\relax
\let\sectionlabel=\relax
\newtoks\chapterstyle        \chapterstyle={\Number}
\newtoks\sectionstyle        \sectionstyle={\chapterlabel\Number}
\newskip\chapterskip         \chapterskip=\bigskipamount
\newskip\sectionskip         \sectionskip=\medskipamount
\newskip\headskip            \headskip=8pt plus 3pt minus 3pt
\newdimen\chapterminspace    \chapterminspace=15pc
\newdimen\sectionminspace    \sectionminspace=10pc
\newdimen\referenceminspace  \referenceminspace=25pc
\def\chapterreset{\global\advance\chapternumber by 1
   \ifnum\equanumber<0 \else\global\equanumber=0\fi
   \sectionnumber=0 \makechapterlabel}
\def\makechapterlabel{\let\sectionlabel=\relax
   \xdef\chapterlabel{\the\chapterstyle{\the\chapternumber}.}}
\def\alphabetic#1{\count255='140 \advance\count255 by #1\char\count255}
\def\Alphabetic#1{\count255='100 \advance\count255 by #1\char\count255}
\def\Roman#1{\uppercase\expandafter{\romannumeral #1}}
\def\roman#1{\romannumeral #1}
\def\Number#1{\number #1}
\def\BLANC#1{}
\def\titlestyle#1{\par\begingroup \interlinepenalty=9999
     \leftskip=0.02\hsize plus 0.23\hsize minus 0.02\hsize
     \rightskip=\leftskip \parfillskip=0pt
     \hyphenpenalty=9000 \exhyphenpenalty=9000
     \tolerance=9999 \pretolerance=9000
     \spaceskip=0.333em \xspaceskip=0.5em
     \iftwelv@\fourteenpoint\else\twelvepoint\fi
   \noindent #1\par\endgroup }
\def\spacecheck#1{\dimen@=\pagegoal\advance\dimen@ by -\pagetotal
   \ifdim\dimen@<#1 \ifdim\dimen@>0pt \vfil\break \fi\fi}
\def\TableOfContentEntry#1#2#3{\relax}
\def\chapter#1{\par \penalty-300 \vskip\chapterskip
   \spacecheck\chapterminspace
   \chapterreset \titlestyle{\chapterlabel\ #1}
   \TableOfContentEntry c\chapterlabel{#1}
   \nobreak\vskip\headskip \penalty 30000
   \wlog{\string\chapter\space \chapterlabel} }

\def\section#1{\par \ifnum\the\lastpenalty=30000\else
   \penalty-200\vskip\sectionskip \spacecheck\sectionminspace\fi
   \global\advance\sectionnumber by 1
   \xdef\sectionlabel{\the\sectionstyle\the\sectionnumber}
   \wlog{\string\section\space \sectionlabel}
   \TableOfContentEntry s\sectionlabel{#1}
   \noindent {\caps\enspace\sectionlabel\quad #1}\par
   \nobreak\vskip\headskip \penalty 30000 }
\def\subsection#1{\par
   \ifnum\the\lastpenalty=30000\else \penalty-100\smallskip \fi
   \noindent\undertext{#1}\enspace \vadjust{\penalty5000}}

\def\undertext#1{\vtop{\hbox{#1}\kern 1pt \hrule}}
\def\ack{\par\penalty-100\medskip \spacecheck\sectionminspace
   \line{\fourteenrm\hfil ACKNOWLEDGEMENTS\hfil}\nobreak\vskip\headskip }
\def\APPENDIX#1#2{\par\penalty-300\vskip\chapterskip
   \spacecheck\chapterminspace \chapterreset \xdef\chapterlabel{#1}
   \titlestyle{APPENDIX #2} \nobreak\vskip\headskip \penalty 30000
   \TableOfContentEntry a{#1}{#2}
   \wlog{\string\Appendix\ \chapterlabel} }
\def\Appendix#1{\APPENDIX{#1}{#1}}
\def\appendix{\APPENDIX{A}{}}
\def\unnumberedchapters{\let\makechapterlabel=\relax \let\chapterlabel=\relax
   \sectionstyle={\BLANC}\let\sectionlabel=\relax \sequentialequations }
%
%
%
\def\eqname#1{\relax \ifnum\equanumber<0
     \xdef#1{{\noexpand\rm(\number-\equanumber)}}%
       \global\advance\equanumber by -1
    \else \global\advance\equanumber by 1
      \xdef#1{{\noexpand\rm(\chapterlabel\number\equanumber)}} \fi #1}
\def\eqinsert#1{\noalign{\dimen@=\prevdepth \nointerlineskip
   \setbox0=\hbox to\displaywidth{\hfil #1}
   \vbox to 0pt{\kern 0.5\baselineskip\hbox{$\!\box0\!$}\vss}
   \prevdepth=\dimen@}}
%

%
%
\def\GENITEM#1;#2{\par \hangafter=0 \hangindent=#1
    \Textindent{$ #2 $}\ignorespaces}
\outer\def\newitem#1=#2;{\gdef#1{\GENITEM #2;}}
\newdimen\itemsize                \itemsize=30pt
\newitem\item=1\itemsize;
\newitem\sitem=1.75\itemsize;     
\newitem\ssitem=2.5\itemsize;     
\outer\def\newlist#1=#2&#3&#4;{\toks0={#2}\toks1={#3}%
   \count255=\escapechar \escapechar=-1
   \alloc@0\list\countdef\insc@unt\listcount     \listcount=0
   \edef#1{\par
      \countdef\listcount=\the\allocationnumber
      \advance\listcount by 1
      \hangafter=0 \hangindent=#4
      \Textindent{\the\toks0{\listcount}\the\toks1}}
   \expandafter\expandafter\expandafter
    \edef\c@t#1{begin}{\par
      \countdef\listcount=\the\allocationnumber \listcount=1
      \hangafter=0 \hangindent=#4
      \Textindent{\the\toks0{\listcount}\the\toks1}}
   \expandafter\expandafter\expandafter
    \edef\c@t#1{con}{\par \hangafter=0 \hangindent=#4 \noindent}
   \escapechar=\count255}
\def\c@t#1#2{\csname\string#1#2\endcsname}
\newlist\point=\Number&.&1.0\itemsize;
\newlist\subpoint=(\alphabetic&)&1.75\itemsize;
\newlist\subsubpoint=(\roman&)&2.5\itemsize;
%

%
%
%
%
\newcount\referencecount     \referencecount=0
\newcount\lastrefsbegincount \lastrefsbegincount=0
\newif\ifreferenceopen       \newwrite\referencewrite
\newif\ifrw@trailer
\newdimen\refindent     \refindent=30pt
\def\NPrefmark#1{\attach{\scriptscriptstyle [ #1 ] }}
\let\PRrefmark=\attach
\def\refmark#1{\relax\ifPhysRev\PRrefmark{#1}\else\NPrefmark{#1}\fi}
\def\refend@{\refmark{\number\referencecount}}
\def\refend{\refend@{}\space }
\def\refsend{\refmark{\count255=\referencecount
   \advance\count255 by-\lastrefsbegincount
   \ifcase\count255 \number\referencecount
   \or \number\lastrefsbegincount,\number\referencecount
   \else \number\lastrefsbegincount-\number\referencecount \fi}\space }
\def\refitem#1{\par \hangafter=0 \hangindent=\refindent \Textindent{#1}}
\def\Ref{\rw@trailertrue\REF}
\def\ref{\Ref\?}

\def\REF#1{\r@fstart{#1}%
   \rw@begin{\the\referencecount.}\rw@end}
\def\REFS#1{\r@fstart{#1}%
   \lastrefsbegincount=\referencecount
   \rw@begin{\the\referencecount.}\rw@end}
\def\r@fstart#1{\chardef\rw@write=\referencewrite \let\rw@ending=\refend@
   \ifreferenceopen \else \global\referenceopentrue
   \immediate\openout\referencewrite=referenc.txa
   \toks0={\catcode`\^^M=10}\immediate\write\rw@write{\the\toks0} \fi
   \global\advance\referencecount by 1 \xdef#1{\the\referencecount}}
{\catcode`\^^M=\active %
 \gdef\rw@begin#1{\immediate\write\rw@write{\noexpand\refitem{#1}}%
   \begingroup \catcode`\^^M=\active \let^^M=\relax}%
 \gdef\rw@end#1{\rw@@end #1^^M\rw@terminate \endgroup%
   \ifrw@trailer\rw@ending\global\rw@trailerfalse\fi }%
 \gdef\rw@@end#1^^M{\toks0={#1}\immediate\write\rw@write{\the\toks0}%
   \futurelet\n@xt\rw@test}%
 \gdef\rw@test{\ifx\n@xt\rw@terminate \let\n@xt=\relax%
       \else \let\n@xt=\rw@@end \fi \n@xt}%
}
\let\rw@ending=\relax
\let\rw@terminate=\relax
\let\splitout=\relax
\def\Closeout#1{\toks0={\catcode`\^^M=5}\immediate\write#1{\the\toks0}%
   \immediate\closeout#1}
%
%
\newcount\figurecount     \figurecount=0
\newcount\tablecount      \tablecount=0
\newif\iffigureopen       \newwrite\figurewrite
\newif\iftableopen        \newwrite\tablewrite
\def\FIG#1{\f@gstart{#1}%
   \rw@begin{\the\figurecount)}\rw@end}

\def\Fig{\rw@trailertrue\def\rw@ending{Fig.~\?}\FIG\?}
\def\fig{\rw@trailertrue\def\rw@ending{fig.~\?}\FIG\?}
\def\TABLE#1{\T@Bstart{#1}%
   \rw@begin{\the\tableecount:}\rw@end}
\def\Table{\rw@trailertrue\def\rw@ending{Table~\?}\TABLE\?}
\def\f@gstart#1{\chardef\rw@write=\figurewrite
   \iffigureopen \else \global\figureopentrue
   \immediate\openout\figurewrite=figures.txa
   \toks0={\catcode`\^^M=10}\immediate\write\rw@write{\the\toks0} \fi
   \global\advance\figurecount by 1 \xdef#1{\the\figurecount}}
\def\T@Bstart#1{\chardef\rw@write=\tablewrite
   \iftableopen \else \global\tableopentrue
   \immediate\openout\tablewrite=tables.txa
   \toks0={\catcode`\^^M=10}\immediate\write\rw@write{\the\toks0} \fi
   \global\advance\tablecount by 1 \xdef#1{\the\tablecount}}
%

%
%
%
\def\getfigure#1{\global\advance\figurecount by 1
   \xdef#1{\the\figurecount}\count255=\escapechar \escapechar=-1
   \edef\n@xt{\noexpand\g@tfigure\csname\string#1Body\endcsname}%
   \escapechar=\count255 \n@xt }
\def\g@tfigure#1#2 {\errhelp=\disabledfigures \let#1=\relax
   \errmessage{\string\getfigure\space disabled}}
\newhelp\disabledfigures{ Empty figure of zero size assumed.}
\def\figinsert#1{\midinsert\Tenpoint\medskip
   \count255=\escapechar \escapechar=-1
   \edef\n@xt{\csname\string#1Body\endcsname}
   \escapechar=\count255 \centerline{\n@xt}
   \bigskip\narrower\narrower
   \noindent{\it Figure}~#1.\quad }
%
%
%
\def\masterreset{\global\pagenumber=1 \global\chapternumber=0
   \global\equanumber=0 \global\sectionnumber=0
   \global\referencecount=0 \global\figurecount=0 \global\tablecount=0 }
\def\FRONTPAGE{\ifvoid255\else\vfill\penalty-20000\fi
      \masterreset\global\frontpagetrue
      \global\lastf@@t=0 \global\footsymbolcount=0}

\def\paperstyle{\letterstylefalse\normalspace\papersize}

%
\def\papersize{\hsize=6.5truein \vsize=9truein
               \skip\footins=\bigskipamount}

\paperstyle   
%
%
%
%
\newskip\frontpageskip
\newtoks\Pubnum
\newtoks\pubtype
\newif\ifp@bblock  \p@bblocktrue
\def\PH@SR@V{\baselineskip=19pt
             }
\def\PHYSREV{\paperstyle\PhysRevtrue\PH@SR@V}
\def\titlepage{\FRONTPAGE\paperstyle\ifPhysRev\PH@SR@V\fi
   \ifp@bblock\p@bblock \else\hrule height\z@ \relax \fi }
\def\nopubblock{\p@bblockfalse}
\def\endpage{\vfil\break}
\frontpageskip=12pt plus .5fil minus 2pt
\Pubnum={}
\def\p@bblock{\begingroup \tabskip=\hsize minus \hsize
   \baselineskip=1.5\ht\strutbox \topspace-2\baselineskip
   \halign to\hsize{\strut ##\hfil\tabskip=0pt\crcr
       \the\Pubnum\crcr\the\date\crcr\the\pubtype\crcr}\endgroup}
\def\title#1{\vskip\frontpageskip \titlestyle{#1} \vskip\headskip }
\def\author#1{\vskip\frontpageskip\titlestyle{\twelvecp #1}\nobreak}

\def\address#1{\par\kern 5pt\titlestyle{\twelvepoint\it #1}}
\def\andaddress{\par\kern 5pt \centerline{\sl and} \address}

\def\abstract{\par\dimen@=\prevdepth \hrule height\z@ \prevdepth=\dimen@
   \vskip\frontpageskip\centerline{\fourteenrm ABSTRACT}\vskip\headskip }

%
%
%

\def\\{\relax \ifmmode \backslash \else {\tt\char`\\}\fi }
\def\sequentialequations{\relax\if\equanumber<0\else\global\equanumber=-1\fi}

\def\journal#1&#2(#3){\unskip, \sl #1\unskip~\bf\ignorespaces #2\rm (19#3),}

\def\topspace{\hrule height 0pt depth 0pt \vskip}
\def\frac#1#2{{\textstyle{#1\over #2}}}

\def\Buildrel#1\under#2{\mathrel{\mathop{#2}\limits_{#1}}}
\def\becomes#1{\mathchoice{\becomes@\scriptstyle{#1}}{\becomes@\scriptstyle
   {#1}}{\becomes@\scriptscriptstyle{#1}}{\becomes@\scriptscriptstyle{#1}}}
\def\becomes@#1#2{\mathrel{\setbox0=\hbox{$\m@th #1{\,#2\,}$}%
	\mathop{\hbox to \wd0 {\rightarrowfill}}\limits_{#2}}}

\let\int=\intop         
\def\lsim{\mathrel{\mathpalette\@versim<}}
\def\gsim{\mathrel{\mathpalette\@versim>}}
\def\@versim#1#2{\vcenter{\offinterlineskip
	\ialign{$\m@th#1\hfil##\hfil$\crcr#2\crcr\sim\crcr } }}
\def\big#1{{\hbox{$\left#1\vbox to 0.85\b@gheight{}\right.\n@space$}}}
\def\Big#1{{\hbox{$\left#1\vbox to 1.15\b@gheight{}\right.\n@space$}}}
\def\bigg#1{{\hbox{$\left#1\vbox to 1.45\b@gheight{}\right.\n@space$}}}
\def\Bigg#1{{\hbox{$\left#1\vbox to 1.75\b@gheight{}\right.\n@space$}}}
%
%
%
\let\sec@nt=\sec
\def\sec{\relax\ifmmode\let\n@xt=\sec@nt\else\let\n@xt\section\fi\n@xt}
\def\obsolete#1{\message{Macro \string #1 is obsolete.}}
\def\firstsec#1{\obsolete\firstsec \section{#1}}
\def\firstsubsec#1{\obsolete\firstsubsec \subsection{#1}}
\def\thispage#1{\obsolete\thispage \global\pagenumber=#1\frontpagefalse}
\def\thischapter#1{\obsolete\thischapter \global\chapternumber=#1}
\def\REFSCON{\obsolete\REFSCON\REF}
\def\splitout{\obsolete\splitout\relax}
\def\prop{\obsolete\prop \propto }
\def\nextequation#1{\obsolete\nextequation \global\equanumber=#1
   \ifnum\the\equanumber>0 \global\advance\equanumber by 1 \fi}
\def\BOXITEM{\afterassigment\B@XITEM\setbox0=}
\def\B@XITEM{\par\hangindent\wd0 \noindent\box0 }
\def\phyzzx{PHY\setbox0=\hbox{Z}\copy0 \kern-0.5\wd0 \box0 X}
%
%
\everyjob{\xdef\today{\monthname\ \number\day, \number\year}}

\catcode`\@=12 
\message{ by V.K.}
        
%


\overfullrule=0pt
\PHYSREV
\normalspace
\vskip 0.6cm
\date={                          }    
\pubtype={Preprint \# 428            }
\titlepage
\title{\bf              Infrared Conductivity of Cuprate Metals:
Detailed Fit Using Luttinger Liquid Theory } 

\author{\bf P.W. ANDERSON}
\address{
Joseph Henry Laboratories of Physics\break
Jadwin Hall, Princeton University \break
Princeton, NJ 08544\footnote{*}{ This work was supported by the NSF, Grant \#
DMR-9104873}
}
\abstract
{Measurements of infrared conductivity in the normal state of the
cuprate layer metals show a characteristic behavior in the plane
of the layers which is in essential agreement among many
experiments.  A simple parametrization of this behavior, proposed
originally by Collins and Schlesinger, and exploited by N.
Bontemps and her group, which gives an adequate
fit over frequencies from a few hundred cm$^{-1}$ to $>5000 $
cm$^{-1}$, is that the phase angle of the complex conductivity
is independent of frequency.  This fit is shown to be a natural
consequence of Luttinger Liquid theory with charge-spin
separation, and determines the exponent of the singularity at the
Fermi surface to be $\sim .15\pm .05$.}
\endpage
\pageno=1
\baselineskip=20pt

The infrared conductivity of the high-$T_c$ cuprates in the
normal state has a characteristic deviation from the normal
``Drude'' behavior of metals, which has sometimes been described
as an additional, distinct ``mid-infrared absorption'' and
sometimes as an extended tail of the low-frequency peak.
Schlesinger,\Ref\one{Z. Schlesinger and R. Collins, Phys. Rev.
Lett., {\bf 65}, 801 (1990)} some years ago, analyzed his data on
the reflectivity of single crystals of $YBCO_7$ in terms of the
conventional expression
$$\sigma={ne^2\over m(i\omega+{1\over\tau})}\eqno(1)$$
with frequency-dependent parameters $m(\omega)$ and
$1/\tau(\omega)$, which showed remarkably simple behavior (see
Fig. 1): $1/\tau$ is proportional to $\omega$, and $m$ has a
slow, approximately logarithmic variation.  There is in fact
little difference in the data among actual experiments, as
opposed to interpretations, on good materials, so we may take
Fig. 1 as typical of optimally doped cuprates, since it is in
essence a heuristic description of the data.

N. Bontemps and collaborators\Ref\two{A. El Azrak, R. Nahoon, A.C.
Boccara, N. Bontemps, M.
Guilloux-Viry, C. Thiret, A. Perrin, Z.Z. Li, H. Raffy, Journal
of  Alloys and Compounds, {\rm\bf 195}, 663 (1993); J. Orenstein,
Phys. Rev. {\bf B49}, 9846 (1995)} have used a similar plot
to describe data over a wide range of frequencies, up to around
8000 cm$^{-1}$, using transmission and reflection data on films of a number of
cuprates, most but not all closely related to YBCO (see Figs. 2
and 3.)  With this wide frequency range the family resemblance of
all of the data becomes striking, particularly plotted using
Schlesinger's parameters.  I believe that there would be little
disagreement as to the general characteristics of the actual data
among these and other experimentalists, except that less highly
doped YBCO samples show ``spin gap'' deviations at the lower end
of the range $(\le 500 {\rm cm}^{-1})$,\Ref\three{G. Thomas, et
al, Phys. Rev. {\bf B42} 6342 (1990)} and that quite impure samples may have a
small residual
resistivity.

We describe a detailed fit to the data of figs.\ 2 and 3  using the
``Luttinger liquid'' hypothesis for the electronic state of the
2D normal metal.\Ref\four{P.W. Anderson, Chapter VI of
forthcoming book; The Theory of Superconductivity; P.W. Anderson,
Proceedings of the ``Materials and Mechanisms of Superconductivity,
High Temperature Superconductors III'', Kanazawa, July 22-26, 1991,
Part I, (eds. M. Tachiki, Y. Muto and Y. Syono), (North-Holland,
Amsterdam, 1991) p. 11,
reprinted from Physica C {\rm\bf 185--189}, 11--16 (1991)}  This result depends
only on rather
general properties of the theory but is totally dependent on its
non-Fermi liquid nature.

The basis of the fit is the remark that the cuprates are in the
``holon non-drag regime'' of Luttinger liquid transport
theory.\refmark{4}  This is the regime where charge excitations
(``holons'') are scattered sufficiently rapidly that they do not
recohere with the spinons after the accelerated electron decays
into charge and spin excitations  (The condition for this regime
is $\omega > (1/\tau)_{\rm holon} >\sim{\omega^2\over E_F}$; the
source of $({1\over\tau_{\rm holon}})$ is probably impurity scattering
at small $\omega$ and phonons at large $\omega$.)

Under these circumstances vertex corrections are damped out by
holon scattering\refmark{9} and the conductivity is given by the
simple one-loop diagram (Fig. 3)
$$\sigma(\omega)\propto {1\over \omega} \int dx\int dt G^e\,(x,t)
G^h\,(x,t)\ e^{i\omega t}\eqno(2)$$
$G^l$ and $G^h$ are the exact (interacting) one-electron Green's
functions for electrons and holes respectively.
The physical process which controls the rate of entropy
production is the decay of the electron and hole into spin and
charge excitations, but this is enabled to act as a resistivity
mechanism by the fact that the momentum decays because the charge
is then scattered by the lattice.  The process is analogous to
phonon scattering in the phonon non-drag regime, where the
momentum decay occurs by the scattering of the phonons by the
lattice which prevents phonon drag, while the entropy production
is caused by phonon emission which is momentum-conserving
and controls the observed resistivity. Ogawa has shown that the
vertex corrections which would invalidate (2) and restore the
Ward identities in a pure sample are cut off by the mean free path
for charge scattering in this regime.  Note that in this regime
neither the the conventional residual resistivity nor phonon
resistivity appear, and they are replaced by the ``linear $T$''
resistivity when the sample is ``impure enough''.

We can evaluate (2) very simply using the fact that $G_1(x,t)$ is
a homogeneous function of $(x,t)$ considered as a single variable.
This is the consequence of the fact that all excitations have a
finite Fermi velocity.
For the Fermi liquid,
$$G_{FL}\propto
{
e^{ik_Fx}\over x-v_Ft
}\eqno(3)$$
homogeneous of order $(-1)$, while for the 1D Luttinger liquid,
$$G_{LL}\propto
{
 e^{ik_Fx}\over
\sqrt{
  (x-v_st)\,(x-v_c t)}\,(x^2-v_c^2t^2)^{{\alpha\over 2}
     }
}\ \ .\eqno(4)$$
which is homogeneous of order $(-1-\alpha)$.  For the 2D liquid $G$ is
an average of an expression like (3) or (4) over the Fermi surface.
For the Fermi liquid, the relevant $G$ in momentum and frequency
space may be approximated by
$$G(p,\omega)\simeq
{
1\over \hbar\omega-
 \vec{(p-p_f)}
 \cdot\vec{v_F}
}$$
where $p_f$, $v_F$ are at the projection of $\vec p$ on the Fermi
surface along $\vec {v_F}$, $p$ assumed close to the Fermi
surface.  A similar construction for the Luttinger liquid will
give a pair of variables $\Delta p=p-p_F$, $\omega$, in which the
Green's function will again be homogeneous of order
$-(1-\alpha)$, but this function has no simple formal expression.
Nonetheless we may in general write, as the appropriate law for
scaling of the low-frequency excitation spectrum,
$$\eqalignno{
G\,(x,t)  &  ={1\over t^{1+\alpha}}\ F\ \Big ({x\over v_Ft}\Big )\ \ ({\rm
						L.L.})&(5)\cr
G\, (p,\omega)  &  ={1\over\omega^{1-\alpha}}\ F\ \Big (
				{(p-p_F)\,v_F\over\omega}\Big)&(6)\cr}$$
where $F$ will depend on the parameters $v_c/v_F$, $v_s/v_F$ as
functions of position on the Fermi surface.  (5) reduces to the
Fermi liquid expression (3) if $\alpha\to 0$.
By a simple scaling argument, we find
$$\sigma(\omega)=
{{\rm const}\over (i\omega)^{1-2\alpha}}
\eqno(7)$$
(7) holds up to an upper frequency cutoff $\Omega=\wedge/\hbar$
of the order of the electron band width $\wedge$.  The sum rule
on conductivity will be satisfied if the coefficient in (7) is
set so that
$$\sigma(\omega)=
{
ne^2\over i\omega m_0
}\
\Big ({
i\omega\over\Omega
 }\Big )^{2\alpha}\ {2\alpha\over \sin \pi\alpha}\eqno(8)$$
Here $m_0$ is the sum rule mass,
$$\int \sigma(\omega)d\omega ={ne^2\over m_0}$$
which should be not far from the band mass: (8) contains all
intraband mass renormalization effects.

I would remind the reader that for the Fermi liquid the integral
(2) is not convergent without a finite lifetime or giving an
imaginary part $\hbar/\tau$ to the energy denominator in $G$.
This gives the characteristic ``Drude'' behavior of ordinary
metals, with $\sigma$ falling off as $1/\omega^2$ at high
frequencies.  The Luttinger liquid is qualitatively different
from a Fermi liquid with small $Z$.

(8) contains only two free parameters, $n/m_0$ and $\alpha$ (the
upper cutoff $\Omega$ merely scales $m_0$ and is not
independent.)
Neither can vary much: $m_0$ must not be much bigger than the
band mass, and $c$-axis Hall data\Ref\five{N.P. Ong, et al,
Phys. Rev. {\bf B46}, 14293 (1992)}
among others tell us
that $n$ is the conventional band filling $\propto 1-\delta$.
$\alpha$ for the 1D Hubbard model is $\le 1/8$, but models with
$\alpha> 1/8$ exist.  There is no fundamental theory of $\alpha $
in 2D.  Vague indications from gauge theory\Ref\six{Z. Zou,
private communication} suggest
$1/6 (2\alpha=1/3)$, while the tomographic picture\Ref\seven{P.W.
Anderson, Chapter VI of forthcoming book;
P.W. Anderson and Y. Ren,
The Normal State of High Tc
Superconductors: a New quantum Liquid,
Proceedings of the
International Conference on  The Physics of Highly Correlated
Electron Systems, Los Alamos,
$\rm\underline {High\ Temperature\
Superconductivity\ Proceedings}$, K. Bedell et al, eds.,
(Addison-Wesley 1990) pp. 3--33} might
suggest agreement with 1D.

The data give two independent measures of $2\alpha$, one from the
slope of $1/\tau$ vs. $\omega$ and one from the dependence of $m$
on $\omega$.  These two numbers are unrelated in the ``marginal
Fermi liquid'' theory\Ref\eight{P. Littlewood and C.M. Varma,
JAP, {\bf 69}, 4979 (1991) }, and their agreement argues
against that theory, as well as does the relatively large value
of $2\alpha$ we find.  The slope of $1/\tau$ (dashed line in Fig. 2a) is
$\sim .7\pm .1$ which gives $\alpha\simeq .15\pm .05$.  The median
slope is used for the dashed line in Fig. 2b, which as you can
see is an adequate fit, although the power-law form is not much
constrained by the data.  On the other hand, the analytic
properties of $\sigma$ require that if it really has constant phase angle
(as Fig. \ 2a shows) it must be a power of $(i\omega)$ (or
logarithmic in the limit $\alpha=0+)$.

Let us summarize the achievements of the Luttinger liquid
hypothesis, coupled with the concept of the holon-non-drag
regime.  (A second way to think of this regime is as one in which
the transport is by spinons\Ref\nine{M. Ogata and P.W. Anderson,
Phys. Rev. Lett., {\bf 70}, 3087, (1993)} and is relaxed by holon
emission and reabsorption).  The original motivation which was
satisfied by this idea was to explain the absence of phonon
scattering effects or, in most cases, of residual impurity
scattering, both of which should be large in most of these
materials.  Let it be explicit that the separation of charge and
spin, though it fails to appear in the formal expression (5) or
(6), which depends only on the ``Fermi surface'' exponent,
$\alpha$, is essential to the entire theory, because of the
concept of ``holon drag''.

Now we see that the theory leads to a unique scaling form for the
conductivity which holds over almost 2 decades of frequency and
for a number of cuprates.  Particularly important, in may view,
is the fact that the expression scales from $>5000 {\rm cm}^{-1}$ to
$<500 {\rm cm}^{-1}$, a property which no alternative theory
motivates in any natural way.

It is interesting that other groups (especially
Bozovic\Ref\ten{I. Bozovic, Recent paper --- Quoted in M.V. Klein}) see
indications of similar behavior in the ``mid-infrared
conductivity''
of a number of other materials, mostly those with other
symptoms of strong correlation phenomena.  With considerable
caution because of the existence of other transport regimes, we
would consider a Luttinger liquid explanation for some of these
cases.
\bigskip

\ack
I would like to acknowledge especially extensive discussions with
Nicole Bontemps, as well as the use of her data.  I also was
stimulated by discussions with R. Laughlin and E. Abrahams.
\endpage
\centerline{\bf FIGURE CAPTIONS}
\medskip

\item{(1)}  Schlesinger's original data on the IR spectrum of
YBCO.  This is repeated in Fig. (2) as the crossed square points. (Ref. (1)).
\medskip

\item{(2)} (a) The Bontemps group's data on a group of cuprates
from Ref. (2): $1/\tau$ vs $\omega$.  Further, more recent data
are given in (2b) and reported in Ref. (11).\Ref\eleven{C.
Baraduc, A. El Azrak, N. Bontemps, J. of Superconductivity, {\bf 8}, 1 (1995)}
\medskip

\item{(2b)} (a)+(b) Same set of data, $m^*$ vs. $\omega$.
\medskip

\item{(3)} Primitive diagram for the conductivity.  Vertex
corrections are omitted for reasons given in the text.
\vfill\eject

\par\penalty-400\vskip\chapterskip\spacecheck\referenceminspace
   \ifreferenceopen \Closeout\referencewrite \referenceopenfalse \fi
   \line{\fourteenrm\hfil REFERENCES\hfil}\vskip\headskip
   \input referenc.txa

\end